**Data-driven sequential analysis of tipping in high-dimensional complex systems**

Tomomasa Hirose[1], Yohei Sawada[2]

Corresponding author: T. Hirose, Department of Civil Engineering, Graduate School of Engineering, the University of Tokyo, Tokyo, Japan, 7-3-1, Hongo, Bunkyo-ku, Tokyo, Japan, hirose-tomomasa693@g.ecc.u-tokyo.ac.jp

---

[1] Department of Civil Engineering, Graduate School of Engineering, the University of Tokyo, Tokyo, Japan

[2] Department of Civil Engineering, Graduate School of Engineering, the University of Tokyo, Tokyo, Japan

## Abstract

Abrupt transitions ("tipping") in nonlinear dynamical systems are often accompanied by changes in the geometry of the attracting set, but quantifying such changes from partial and noisy observations in high-dimensional systems remains challenging. We address this problem with a sequential diagnostic framework, Data Assimilation-High dimensional Attractor's Structural Complexity (DA-HASC). First, this method reconstructs system's high-dimensional state using data assimilation from limited and noisy observations. Second, we quantify a structural complexity of the high-dimensional system dynamics from the reconstructed state by manifold learning. Third, we capture underlying changes in the system by splitting the reconstructed time-series into sliding windows and analyzing the changes in the temporally local attractor's structural complexity. The structural information is provided as graph Laplacian and measured by Von Neumann entropy in this framework. We evaluate DA-HASC on both synthetic and real-world datasets and demonstrate that it can detect tipping under high-dimensionality and imperfect system knowledge. We further discuss how this framework behaves across different tipping mechanisms.

## Lead paragraph

Tipping points—abrupt shifts between dynamical regimes—can arise in many nonlinear systems including the Earth system, yet the relevant changes are not always captured by an indicator computed from a single scalar time-series. This study proposes a practical way to monitor such structural change in high-dimensional systems from imperfect observations by combining state reconstruction and a geometry-based complexity measure: we map short segments of trajectory data to a graph and quantify its spectral complexity using an entropy derived from the graph Laplacian. Tracking this quantity over time reveals when the underlying dynamics reorganize by capturing changes in the effective degrees of freedom of the observed state-space geometry. We illustrate the method on benchmark models, discuss how its behavior differs across tipping mechanisms, and demonstrate its practicality in high-dimensional applications.

# 1.Introduction

Abrupt transition points ("tipping points"), critical thresholds beyond which a system experiences a rapid and often irreversible shift, are a central theme in nonlinear dynamics and complex systems. In climate science, tipping has been highlighted in the context of climate change over the last two decades (van Nes et al., 2016), with widely discussed examples including melting of Greenland Icesheet, dieback of Amazon Rainforest, collapse of Atlantic Meridional Overturning Circulation (AMOC) etc. (Lenton et al., 2008). It is of paramount importance to develop mathematical methods to realize an improved understanding of these tipping phenomena and apply them to construct better Early Warning Signals (EWS).

Most of the widely used early warnings are rooted in critical slowing down (CSD) and are primarily designed and used for systems that gradually approach a bifurcation, i.e., bifurcation-induced tipping (B-tipping) (Scheffer et al., 2009; Boers, 2021). Two limitations are particularly explicit (Dakos et al., 2024). First, generic CSD-based warnings are not specific to abrupt and irreversible transitions; they can also respond to smooth and reversible changes, which induces false positives. Second, many early-warning tools are well tailored to unidimensional observables (Dakos et al., 2008), whereas most real systems are high-dimensional. The researches have thus increasingly emphasized approaches that treat multivariate dynamics explicitly via evolving networks analysis (Lu et al., 2021; Zhang et al., 2024) or dimension reduction by Principal Component Analysis (PCA) also known as Empirical Orthogonal Function (EOF) in climatology (Held & Kleinen, 2004), or to combine generic and system-specific information in a principled way (Flores et al., 2024).

However, these multivariate and hybrid directions introduce practical trade-offs. Network-based early warnings can be highly sensitive to how the network is defined, to thresholding choices, and to the stability of inferred connections. Stabilizing such pipelines typically requires phenomenon-informed design choices. PCA/EOF based approaches face a closely related issue: the extracted dominant modes and their trends can vary across parameter choices and over time. Therefore, even with careful preprocessing, their consistent effectiveness may break down under nonlinear, regime-dependent responses manifested through interacting drivers and delayed adjustments without system-specific prior knowledge about the dynamics. While Machine-Learning approaches (ML) have been proposed as an alternative approach and can achieve high accuracy on benchmark datasets (Bury et al., 2023), they make this 'dependency on design choices problem' even more explicit. Their performance depends critically and essentially on the construction of the training corpus and its match to the target system (Dakos et al., 2024).

These challenges become even more acute when focusing on applying to real Earth-system datasets. In these settings, observational sparsity, strong noise, and changes in the observation system can substantially affect early-warning estimates. Data assimilation (DA) offers a principled way to address these issues by integrating observations into process-based models to produce consistent state estimates and uncertainty-aware ensembles (Carrassi et al., 2018), and existing datasets are usually reanalysis data after DA. In this context, DA should be treated as an explicit component of the tipping-detection framework rather than as a "pre-preprocessing".

These considerations motivate a tipping indicator that is training-free, multivariate, and operationally robust under noisy, sparse and indirect observations. We therefore frame tipping analysis as a trajectory and geometry structure-aware monitoring problem, where tipping-relevant change is reflected in the observed or assimilated (high-dimensional) state space rather than univariate temporal statistics. This viewpoint is related to orbit-based diagnostics from nonlinear dynamics. Most popularly, the Largest Lyapunov Exponent (LLE) quantifies asymptotic dispersion of deterministic chaotic attractor orbits (Benettin et al., 1980). In stochastic or partially observed systems, Ensemble-Averaged Pairwise Distance (EAPD) function has been proposed as a practical surrogate ("Instantaneous" LLE), with recent studies indicating potential applicability to high-dimensional systems (Jánosi & Tél, 2024).

First, we use DA as a key component to reconstruct the underlying high-dimensional system from limited data. Second, instead of applying conventional dimension reduction methods, we adopt manifold-learning-based approach to extract structural information without cutting off important high-dimensional information. We quantify this structural complexity as entropy, interpreting its fluctuations as a signal of regime shifts. This indicator is thus designed to capture not only the complexity of stable phases before and after a tipping event, but also the structural change during the tipping process itself. As a practical application, we partition the time series into sliding windows, calculate values for each window, and tracked the changes. Here, we term this indicator High-dimensional Attractor's Structural Complexity (HASC). The integrated analysis framework of DA and HASC is specifically called DA-HASC.

This paper is organized as follows. Section 2 introduces the proposed DA-HASC framework and its theoretical backgrounds as a quantification method of the dynamical system's structural complexity. Section 3 demonstrates that HASC is a consistent and practically robust indicator of structural complexity of the systems as well as chaos. Section 4 further explores the use of proposed indicator as a tipping indicator following the theories on the tipping phenomena. Section 5 summarizes the paper.

## 2.Method

DA-HASC consists of 3 steps: DA, structural information extraction, and complexity evaluation. These 3 steps are explained in the following Sections 2.1-2.3. Then, we will discuss the practical implementation and hyperparameter selection in Section 2.4..

### 2.1. Data Assimilation (DA)

DA combines the prior knowledge of a system's numerical dynamics and real-world observations to obtain the best possible state estimation that explains real-world well. In DA, the underlying discrete-time dynamical system is formulated as:

$$X(t) = f(X(t-1), \theta(t-1), u(t-1)) + q(t-1) \tag{1}$$

where $X(t), \theta(t), u(t),$ and $q(t)$ denote state variables, model parameters, external forcing, noise process at timestep $t$, respectively. Observation for this discrete dynamical system can be written as:

$$Y(t) = h(X(t)) + r(t) \tag{2}$$

where $Y(t),$ and $r(t)$ denote observation and noise process, respectively. In this paper, we used Local Ensemble Transform Kalman Filter (LETKF), which has been widely applied for high-dimensional spatiotemporally chaotic systems (Hunt et al., 2007). In (1), by assuming that the system has time-varying parameters, DA is applied so that the following cost function is minimized:

$$J = \frac{1}{2}\begin{pmatrix}X - \bar{X}^b \\ \theta - \bar{\theta}^b\end{pmatrix}^T \begin{pmatrix}B_x & B_{x\theta} \\ B_{\theta x} & B_\theta\end{pmatrix}^{-1} \begin{pmatrix}X - \bar{X}^b \\ \theta - \bar{\theta}^b\end{pmatrix} + \frac{1}{2}(Y - h(X))^T R^{-1}(Y - h(X)) \tag{3}$$

where $\bar{X}^b$ and $\bar{\theta}^b$ denote the forecast ensemble means and $B_x, B_\theta$ are the background error covariances of the state and parameter, and $B_{x\theta}(B_{\theta x})$ represents their cross-covariance, enabling indirect parameter updates through state-parameter correlations. The matrix $R$ is the observation error covariance associated with $Y$ and $h$. This state augmentation form of DA for simultaneous estimation of states and parameters has been widely adopted (e.g., (Sawada & Duc, 2024)).

## 2.2. Structural information extraction

The fundamental assumption of manifold-based dimension reduction (DR) is that high-dimensional data plots are located on a lower-dimensional data manifold. While such DR techniques usually aim to visualize data structure, they also offer a way to naturally extract manifold's geometrical features through graph expression by constructing K-nearest neighbor graph (K-NNG). The extracted information is then embedded into low-dimensional space (usually 2 or 3 dimensions) using a graph layout algorithm (Tang et al., 2016). Among several DR methods, Uniform Manifold Approximation and Projection (UMAP) (McInnes et al., 2018) is remarkable for its better preservation of global structure and stronger scalability compared to other popular manifold learning methods such as t-SNE (Maaten & Hinton., 2008) or Isomap (Tenenbaum et al., 2000). As our primary interest lies in detecting change in global structure of data-manifold in high-dimensional spaces, these UMAP features are significantly advantageous.

Although UMAP is grounded in Riemannian geometry and algebraic topology, its algorithm can be understood more intuitively in terms of implementation. In UMAP, the data manifold is considered as a Riemannian manifold (i.e., a smooth manifold where choices of inner products on each tangent space called Riemannian metric are given). Local K-NNGs are constructed by approximating local distances as if they were induced by a Riemannian metric, and by combining them, the topological representation of the Riemannian manifold is generated in discrete graph expression. In visualization process, UMAP uses spectral embedding as initialization of embedding, which is later optimized through minimizing cross entropy between topological representations of actual and embedded layout. For more detailed theoretical and practical steps of UMAP, refer to the original paper (McInnes et al., 2018). Here, we only focused on the process of constructing topological representation and ignore the visualization steps.

Upon using UMAP, there are three key considerations that we want to clarify here. First, UMAP relies on 3 fundamental prerequisites for analyzed data: uniform distribution of data plots over their data manifold, local consistency of Riemannian metrics and local connectivity of manifold. Although it is hard to rigorously see if these conditions are satisfied, McInnes et al. (2018) concluded that real data usually meets these requirements, and even for data with inequivalent data distribution density, there already exists a method to apply UMAP on them (DensMAP; (Narayan et al., 2021)).

Secondly, one should carefully tune the hyperparameter $k$, which indicates the number of neighbors to consider when approximating local metrics. Note that UMAP includes other parameters, but we only focus on $k$ as a key parameter and all other parameters are fixed to their default values. Setting $k$ bigger allows one to see data structure in a broader perspective.

Third, UMAP offers discrete graph expressions in a non-standard form, which is weighted fuzzy undirected adjacency matrix. Let $G = (V, E)$ be an undirected graph with set of vertices $V = \{1, 2, \ldots, n\}$ and set of edges $E$. Usually, adjacency matrix $A(G)$ for undirected n-vertex graph $G(V, E), |V| = n$ is size n symmetric matrix defined as:

$$A_{ij} := \begin{cases} 1, & if \ \{i,j\} \in E; \\ 0, & otherwise. \end{cases} \ (i, j \in V) \tag{4}$$

However, in weighted fuzzy graph, $A_{ij}$ can take any value between 0 and 1, and the weight implies the probability of the edge existence. This representation contains stochastic factor, which is another promising feature in this framework particularly when dealing with data that contains uncertainty.

## 2.3. Complexity evaluation

To quantify the complexity of the obtained graph expression of structural information, various indicators have been proposed in graph theory and network theory (Zenil et al., 2018). Among them, we adopt the Von Neumann Entropy (VNE) of the normalized graph Laplacian in this study. This indicator is theoretically derived from spectral graph theory, which can provide natural connections between graph structure, diffusion processes and spectral properties. Together with insights from semi-classical theory and quantum chaos theory, this measure can offer physically meaningful interpretation of complexity and regime transitions in the perspective of chaotic dynamical system.

The use of VNE as a graph complexity measure has been well explored in graph theory (Liu et al., 2022). VNE was originally introduced in the context of quantum measurement to describe the irreversibility in quantum states. Based on quantum information theory, VNE was extended to graph state via graph density matrices, as in (Braunstein et al., 2006). Given an undirected graph with n-vertices $G(V, E)$, $|V(G)| = n$, degree matrix $D(G)$ can be written as:

$$D_{ij} := \begin{cases} d_G(v_i) & if \ i = j; \\ 0 & if \ i \neq j. \end{cases} \ (i, j \in V) \tag{5}$$

where $d_G(v_i)$ is the number of edges adjacent to a vertex $v_i$. Graph Laplacian can be obtained from $D(G)$ and $A(G)$ by $L(G) = D(G) - A(G)$. The symmetric normalized Laplacian $\mathcal{L}(G)$ and the corresponding graph density matrix can be derived as:

$$\rho_G := \frac{\mathcal{L}(G)}{n}, \mathcal{L}(G) := D(G)^{-\frac{1}{2}} L(G) D(G)^{-\frac{1}{2}} \qquad (6)$$

,which fulfills the requirements for a density matrix: positive semi-definite, Hermitian, and trace-one (Anand et al., 2011). Then, the Von Neumann Entropy (VNE) can be calculated as:

$$S(\rho_G) = -tr(\rho_G \ln \rho_G) = -\sum_i \lambda_i \ln \lambda_i \qquad (7)$$

where spectral decomposition of $\rho_G$ is $\rho_G = \sum_i \lambda_i |\varphi_i\rangle\langle\varphi_i|$. Although complete interpretation or application of VNE as a graph or network complexity indicator is still being studied, many papers, especially from complex network theory, have already made VNE-based analysis (Ye et al., 2018; Vera et al., 2021). This way of formulating VNE from a graph is often referred to as the Braunstein-Ghosh-Severini (BGS) entropy (Braunstein et al., 2006).

This graph Laplacian works as a discrete approximation of Riemannian manifold's Laplace-Beltrami operator (Belkin & Niyogi, 2008), which naturally provides geometric interpretation of our indicator. In this perspective, our approach is deeply connected with Nonlinear Laplacian Spectral Analysis (NLSA) (Giannakis & Majda, 2012), in which Laplace-Beltrami eigenfunctions form an orthogonal basis for $L^2(M,\mu)$, enabling spatiotemporal modes to be extracted via diffusion geometry. Here, in contrast to explicitly identifying dominant modes, we focus on quantifying how the distribution of such modes varies across instantaneous time windows. A low VNE value corresponds to a concentrated spectral distribution, indicating that trajectories are constrained to a limited set of effective directions and exhibit relatively coherent, low-dimensional dynamical evolution. On the other hand, a high VNE value reflects a broader and more heterogeneous spectral distribution, suggesting that trajectories spread across multiple directions and explore the manifold in a more isotropic and structurally complex manner. Interpretation of VNE as an indicator of heterogeneity is also discussed in the context of graph structure by Liu et al. (2022).

From an information-theoretical perspective, the graph density matrix can be interpreted as a spectral representation of connectivity induced by the trajectories. In this context, the eigenvalues $\{\lambda_i\}$ characterize how diffusion-like connectivity is distributed across the data manifold. Then,

VNE quantifies the degree of spectral delocalization, measuring how evenly the spectral components are spread rather than dominated by certain modes. This viewpoint is analogous to ideas in semiclassical and quantum chaos studies (Haake et al., 2010; Wimberger, 2022), where changes in spectral statistics are often associated with transitions between integrable and chaotic dynamical behavior. While no direct correspondence to Hamiltonian spectra is assumed here, this analogy provides an intuitive interpretation of increasing VNE as reflecting enhanced structural complexity of the underlying dynamics.

In short, in a heuristic sense, the derived spectrum of the normalized Laplacian approximates the diffusion spectrum associated with the invariant sampling measure. VNE therefore measures the uncertainty of diffusion modes, analogous to the semiclassical correspondence between chaotic trajectories and spectral delocalization. In this paper, we often refer to this property as "spread of effective dimension" to offer intuitive understandings. These various theoretical connections enrich the interpretability of our indicator, and thus we explore further applications using this quantified structural information.

## 2.4. Implementation

For practical implementation, reconstructed data from given sparse observation and partial knowledge of the system dynamics via DA is used as the input to UMAP. For temporal tipping analysis, the time-series data are segmented into windows of fixed time-length, and then complexity of each temporally local attractor is independently evaluated. From the sparse undirected graph adjacent matrix returned by the $fuzzy\_simplicial\_set$ function in the UMAP implementation (McInnes, 2026), the normalized graph Laplacian matrix is computed. When computing VNE, one must calculate all the eigenvalues of given density matrix, which results in significant computational cost particularly in high-dimensional settings. Thus, we adopt a popular approximation of VNE, quadratic approximation (Passerini & Severini, 2011; Han et al., 2012):

$$S(\rho_G) \approx \frac{tr(\mathcal{L}(G))}{2} - \frac{tr(\mathcal{L}(G)^2)}{4} \qquad (8)$$

Note that there are two common choices for defining VNE on graphs, based on the Laplacian and the normalized Laplacian, and the use of its quadratic approximation has been actively studied. According to network entropy literatures, while the derivation from the normalize Laplacian lacks the assurance of the aggregate subadditivity (De Domenico & Biamonte, 2016), it is effective to show the relative structural change regardless of scale. In our problem settings, scale-invariance for monitoring relative structural changes weighs more than the rigorous interpretation of network aggregation. Furthermore, Minello et al. (2019) evaluated the quality of quadratic approximations for both types of graph VNE and showed that the approximation error depends on network topology. In particular, they reported that the VNE based on normalized Laplacian is better approximated for Erdős–Rényi and scale-free networks with low edge density, which is consistent with our setting since the UMAP k-NN graph is sparse and hub-like by its construction. From these insights, we conclude that the adaptation of this approximation for the normalized Laplacian-based VNE is reasonable especially when analyzing high-dimensional data.

The hyper-parameters for HASC are the window length $w$, the stride $s$, and the UMAP neighborhood size $k$. We denote a setting in the framework as $HASC(w, s, k)$ when describing experiment settings in this paper. Conceptual DA-HASC algorithm flow diagram is Figure 1.

Given time-series data or optionally reanalysis data $(X_1, X_2, \ldots, X_T)$, where $X_t \in \mathbb{R}^d$ denotes the d-dimensional state vector at time $t$, we form overlapping windows

$$W_n = \{X_{t_n}, X_{t_{n+1}}, \ldots, X_{t_n+w-1}\}, t_n = 1 + (n-1)s \tag{9}$$

for $n = 1, \ldots, N$, with $N = \left\lfloor \frac{T-w}{s} \right\rfloor + 1$. Let

$$\mathcal{C}_k(\cdot): (\mathbb{R}^d)^w \to \mathbb{R} \tag{10}$$

denote the window-to-scalar complexity functional induced by the HASC pipeline with neighborhood size $k$. We then define the HASC indicator time series under the setting $HASC(w, s, k)$ by

$$H_n = \mathcal{C}_k(W_n), n = 1, \ldots, N \tag{11}$$

and associate each value $H_n$ with the window centre time $t_n + \frac{w-1}{2}$. $\{H_n\}_{n=1}^N$ is the output used for monitoring structural change over time. In early-warning application sections, we consider a right-aligned (causal) windowing convention to avoid using future information. Using the same evaluation grid, the right-aligned window ending at $t_n$ as

$$W_n^R = (X_{t_n-w+1}, X_{t_n-w+2}, \ldots, X_{t_n}), n = n_0, \ldots, N \tag{12}$$

where $n_0 = \left\lfloor \frac{w}{s} \right\rfloor$. We then compute

$$H_n^R = \mathcal{C}_k(W_n^R), n = n_0, \ldots, N \tag{13}$$

so that $H_n^R$ is the indicator available at time $t_n$ using only information up to $t$.

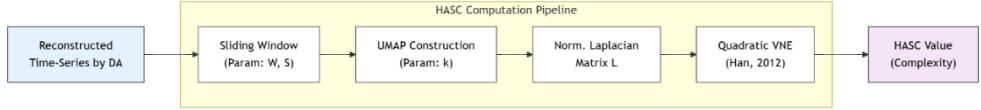

**Figure 1. Flow diagram of DA-HASC algorithm**

## 3. HASC as an indicator of chaos

In this section, we demonstrate that HASC serves as an indicator of chaos and is comparable to LLE but capturing a fundamentally different aspect of the system. Together with DA, it also exhibits strong scalability in high-dimensional systems and practical robustness to noisy observations.

### 3.1. Lorenz63 model: Comparison with LLE

To validate that the HASC indicator accurately expresses structural complexity, we use the Lorenz63 model. The Lorenz63 model introduced by Lorenz (1963) is governed by the following equations:

$$\dot{X} = -\sigma X + \sigma Y, \tag{14}$$

$$\dot{Y} = -XZ + \rho X - Y, \tag{15}$$

$$\dot{Z} = XY - \beta Z \tag{16}$$

where $(\sigma, \rho, \beta)$ are parameters. In the original paper, chaotic behavior was observed with $(\sigma, \rho, \beta)$ = (10,28,8/3). When $\sigma$ and $\beta$ are fixed as 10 and 8/3 respectively, the system undergoes a regime shift from periodic to chaotic attractors as $\rho$ increases approximately beyond 24.74.

To compare HASC with a conventional analytical method, we calculated HASC and instantaneous LLE for each time-window. Thus, we analyze with $HASC(200,10,10)$ and applied the same window length when computing LLE. Estimation of LLE from relatively short time-series data had been studied for decades, and one of the established algorithms is Rosenstein's algorithm (Rosenstein et al., 1993). We employ this algorithm to approximate LLE.

We set two different fixed settings for parameter $\rho = 150$ and $\rho = 28$ with the same initial condition $(1.0, 1.0, 1.0)$. For $\rho = 150$, the trajectory converges to a limit cycle, while for $\rho = 28$, the chaotic butterfly attractor appears. Figure 2 shows that the overall behaviors of HASC and LLE coincides in different orbit geometries, which can be explained by famous Kaplan-Yorke conjecture (Kaplan & Yorke, 1979; Frederickson et al., 1983). The information dimension of an attractor follows the Lyapunov spectrum, and the effective dimension measured as VNE reflects this dynamical feature.

Inspecting correlation between two indicators for chaotic settings yields more rigorous interpretations of the similarity and differences between HASC and LLE. Figure 3 shows that, although total correlation between them is low ($\approx 0.2$), when focusing only on windows inside wings of the attractor, their correlation is significantly high ($\approx 0.7$). This is because, around the saddle point, LLE tends to stay low while HASC value tends to be higher. As shown in Kuptsov & Parlitz (2012), in regions of homoclinic tangencies (typically near saddle points), the stable and unstable manifolds become collinear. This loss of hyperbolicity results in a singularity of covariant Lyapunov vectors (CLV), which cancel out each other and suppress LLE. However, HASC captures this merger of clusters and expresses it as a spike in structural complexity. This induces the strong agreement between HASC and LLE on metastable regime, and at the same time the local disagreement around saddle point.

We briefly show and discuss the parameter sensitivity of HASC analysis on this experiment in the Supplement (S1). Although the rough trend of the results is insensitive to hyperparameters, setting window length and resolution of manifold-learning is not a mere hyperparameter tuning, but a definition of temporal and geometric coarse-graining. Thus, these parameter settings should be well considered based on the dynamics one wants to analyze.

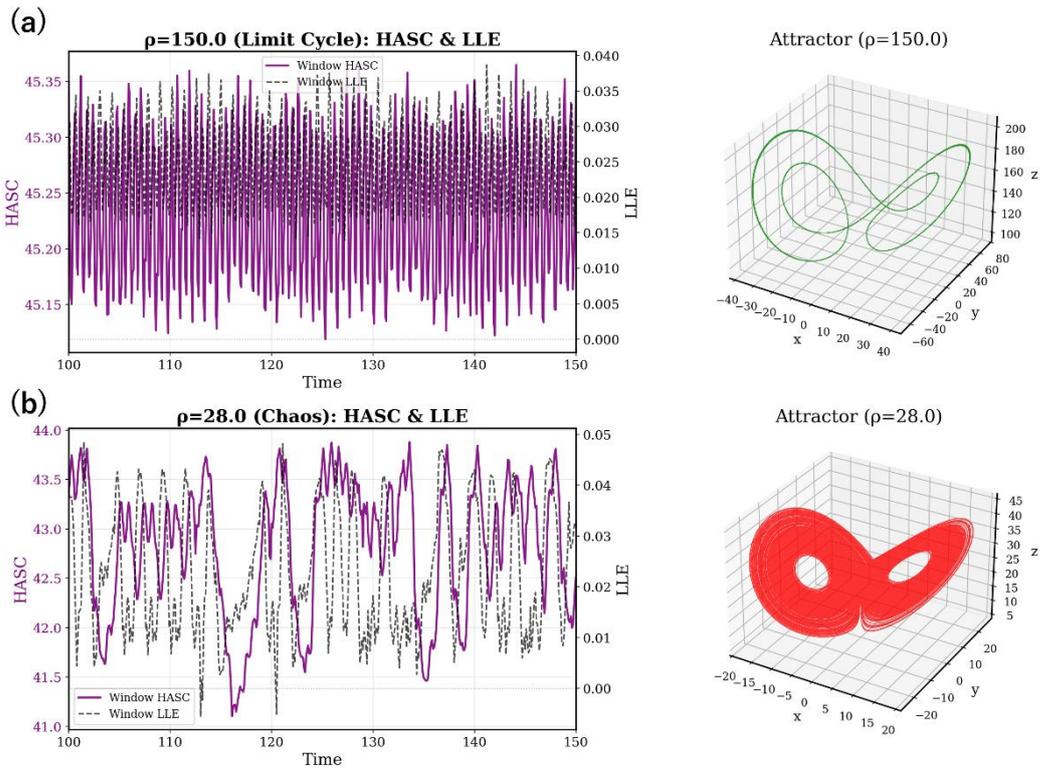

**Figure 2.** Plot of HASC/Instantaneous LLE from t=100-150, and attractors for Lorenz63 model with different parameter settings (a) $\rho = 150$ (b) $\rho = 28$. Purple lines shows the HASC value and black lines shows the estimated instantaneous LLE value.

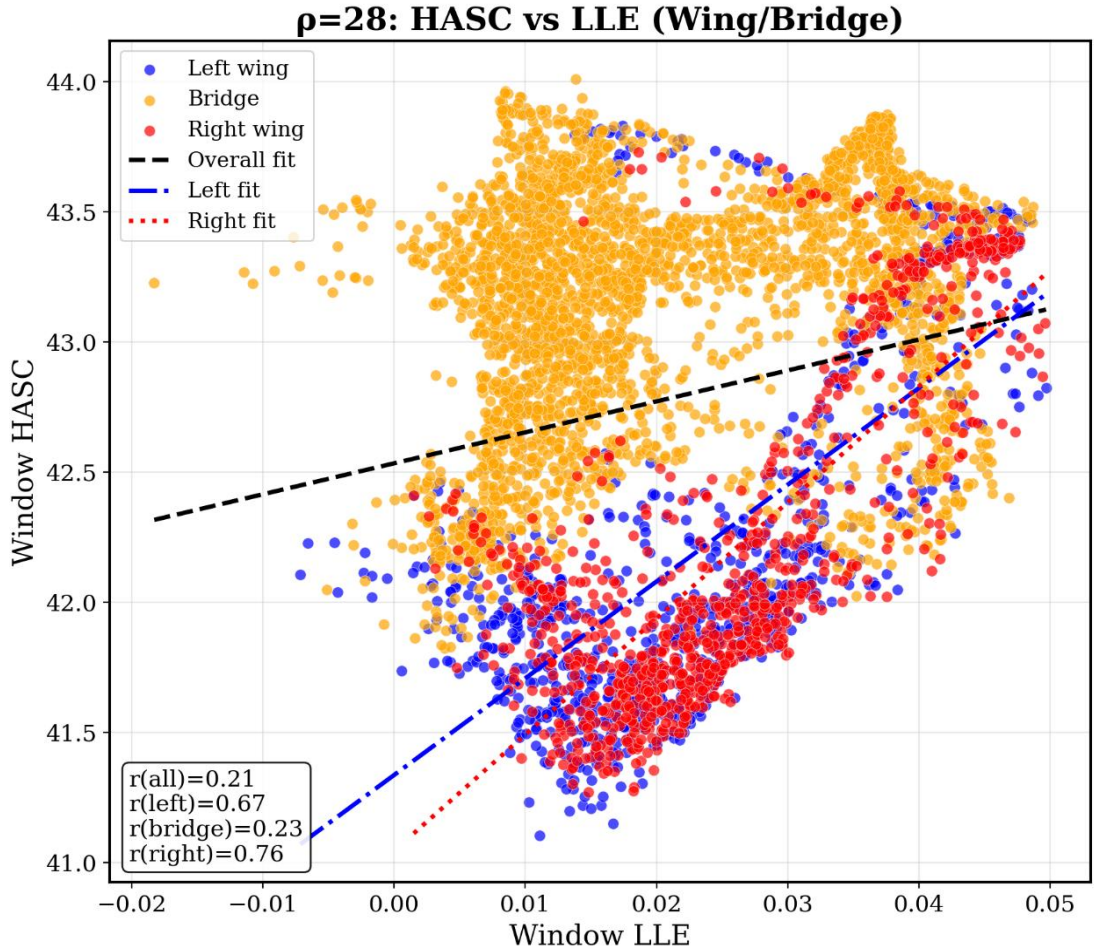

**Figure 3. Scatter plot of HASC and LLE value in the Lorenz63 model experiment with setting $\rho = 28$. Blue dots indicate that the value is calculated in windows fully on left wings of the attractor, while red dots on right wings. Any value except for windows fully on wings of the attractor are denoted as windows on bridge, and indicated as orange dots in the plots. Pearson correlation coefficient and linear fit line for each dot are shown in the figure.**

## 3.2. Robustness of DA-HASC to high-dimensionality and corrupted observations

We show the effectiveness of HASC under a high-dimensional system. Also, the integration of DA into HASC is demonstrated as an effective method against noisy and sparse observations. We used the Lorenz96 model (Lorenz, 2006), which is widely used in studies on DA (Szendro et al., 2009; Miwa & Sawada, 2024; Sawada & Duc, 2024). Here, the Lorenz96 model is formulated as:

$$\dot{x}_j = x_{j-1}(x_{j+1} - x_{j-2}) - x_j + F + 2.0 \cdot \xi_{j,\lfloor t/\Delta T \rfloor}, \quad j = 1, \ldots, n \tag{17}$$

where F is the forcing parameter that controls the degree of chaos in the system. Chaotic behavior was observed for F larger than 8 (Lorenz, 2005). The noise term $\xi_{j,k} \sim \mathcal{N}(0,1)$ was added to the forcing to mimic more realistic features of the real-world climate system.

In order to mimic the tipping behavior described numerically by Gelbrecht et al. (2021), we introduced a time-varying parameter F. Note that the purpose of this setup is not to study the tipping phenomena itself, but to numerically reproduce the tipping-behavior to focus exclusively on the robustness of DA-HASC to high-dimensionality and corrupted observations in this experiment.

The system first stays in the phase with F=8.0 for 36000 timesteps. Then, F linearly increases with strong perturbation from 8.0 to 11.6 over the next 72000 timesteps. Finally, the system stays in the phase with F=11.6 for another 36000 timesteps. See Figure 4 for the parameter setting and the example of the state snapshots from each phase. In this section, we refer to the phase corresponding to each F dynamics, as follows. The first 36000 timesteps are "stable phase", the second linear increase phase is "transient phase" and the last 36000 timesteps are "chaotic phase".

The outputs of this model are considered as the "nature state", from which noisy and sparse observations are generated. LETKF with ensemble size 40 is then applied to the obtained observations and the Lorenz96 model. The Lorenz96 model used in LETKF does not include any knowledge of the change in F. Thus, the nature state and the prediction model used in LETKF diverge during transient and chaotic phases. The assimilated state estimates were then used as input data to $HASC$ (4000,500,50).

As the input of HASC is not the truth state, it is expected that the result of tipping analysis differs from the accuracy of the estimated states. In order to check the practical robustness of DA-HASC, we vary the observation sparsity, observation error, and the accuracy of prior F independently from a baseline configuration. The baseline is set to be 13 observable states (subsampled the grid by a factor of 3), an observation error of 0.1, and a correct prior F of 8.0. Additionally, a "worst-case experiment" is conducted by combining severe conditions (6 observables, error of 2.8, and prior F of 5.0). In total, 11 unique experiments are performed, and the derived VNE time series are compared with the "nature state".

Figure 5 illustrates the results of the baseline and the worst-case experiments, highlighting a decoupling between the accuracy of state estimation and the fidelity of topological geometric construction. Panel (a) displays the time evolution of the RMSE between the nature state and the LETKF estimate. In the baseline experiment ("good"), the RMSE remains low during the stable phase but degrades significantly as the system enters the transient phase. This degeneration is expected as the LETKF uses the model with a fixed forcing parameter whereas the nature run undergoes a forcing increase. The worst-case experiment ("terrible") exhibits an even higher RMSE from the outset due to the sparsity and high noise of observations. This conventionally suggests that the estimated states are unreliable for dynamical analysis. However, the HASC analysis in panel (b) reveals the contrasting perspective, as the VNE time-series computed from the LETKF estimate reproduce the trend in the one computed from the nature state. This demonstrates that, if the knowledge of LETKF on the system dynamics is mostly preserved, the geometry of the attractor is also preserved in the assimilated data manifold even when the individual state trajectories are displaced.

This result suggests that for the purpose of tipping detection, perfect state reconstruction is not a prerequisite. What is crucial is that the underlying dynamics are correctly captured in qualitative way. As long as the DA process preserves this dynamical integrity, from imperfectly observed data, HASC can extract the correct dynamical signals as a trend.

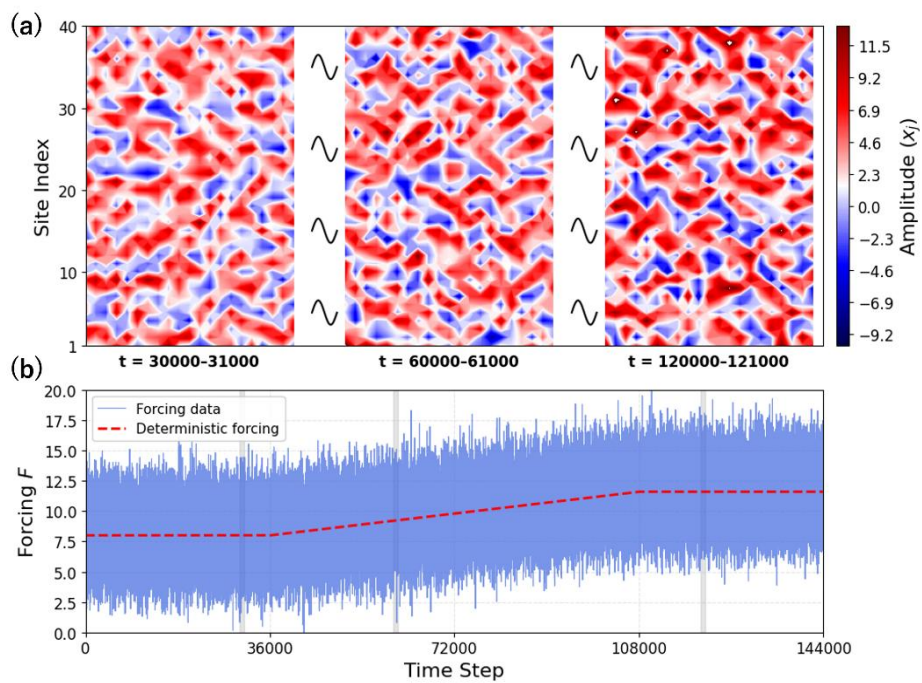

**Figure 4. Experiment setting of Lorenz96 model. (a) Hovmöller diagram of "stable phase",**

"transient phase", and "chaotic phase" of forced Lorenz96 experiment. The horizontal axis represents time, and the vertical axis indicates the dimension index. (b) Time series of the first dimension of the corresponding parameter. The red dotted line indicates the deterministic forcing, while the blue solid line shows the actual forcing data with additive noise. The black shaded region corresponds to the time interval displayed in the top panel.

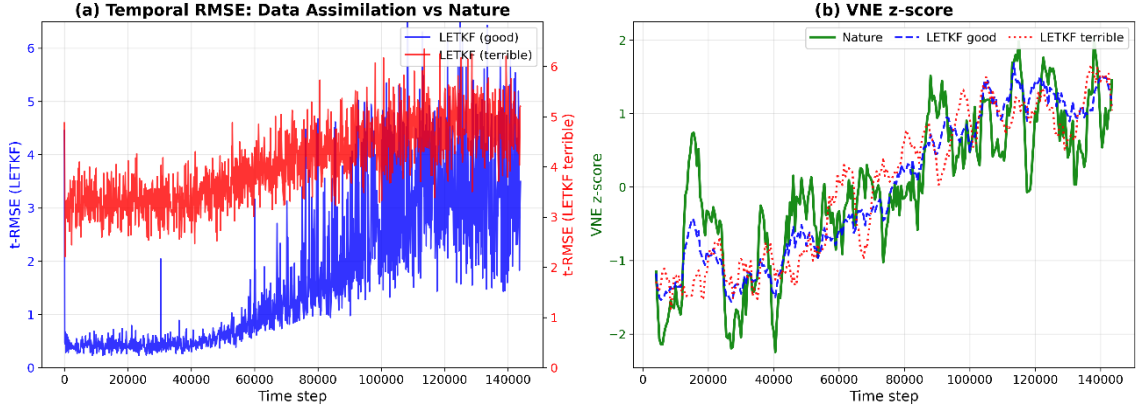

Figure 5. Robustness of DA-HASC against model error and observational degradation. (a) Time evolution of the t-RMSE of the state estimation by LETKF. The blue line represents the baseline "good " condition (13 observables, observation error 0.1, prior F is correct), while the red line represents the "worst-case" (terrible) condition (6 observables, observation error 2.8, prior F is incorrect). (b) Corresponding time evolution of the Z-scored VNE time-series. The green solid line denotes the VNE calculated from the true nature state, and the blue and red dashed lines correspond to the baseline and worst-case experiments, respectively.

## 3.3. Feasibility

Furthermore, to check practical applicability of this algorithm against high dimensional systems, computational cost to compute HASC for each window was evaluated. We generate Gaussian random numbers and evaluate the log-scale metrics while varying a single parameter (model dimension and length of window, up to 8 million).

As expected, HASC computation time is mostly determined by graph construction phase in UMAP, which resulted in $\approx O(N)$ scaling in model dimension and $\approx O(N^{1.1})$ scaling in window length empirically. The K-NNG construction algorithm utilized in UMAP (Dong et al., 2011) reports the window length scaling as $O(N^{1.14})$, which almost fits our results. This $O(D \cdot W^{1.14})$ scaling ($D, W$ denotes the system dimension and the window length) enables HASC to be applicable to high-dimensional simulation results, which is demonstrated later in this paper. Note that setting the k-neighbor parameter lower also contributes to the reduction of computation time.

# 4. HASC as an indicator of tipping

In this section, we discuss the applicability of the proposed HASC as an indicator and an EWS of tipping and demonstrate its practicality through an application to real climatological simulation data. Following the classification of tipping phenomena by Ashwin et al. (2012), we consider three types of tipping: Bifurcation-induced tipping (B-tipping), Noise-induced tipping (N-tipping) and Rate-induced tipping (R-tipping). Theoretically, HASC can serve as an EWS for B-tipping, where tracking a geometrical property known as the "slow manifold" contributes to the tipping prediction. In contrast, for N- and R-tipping, HASC functions primarily as a detection method of structural reconstruction, which can occasionally serve as an imminent EWS. We call HASC as "indicator of tipping" in this paper rather than EWS in this sense.

In systems exhibiting a separation of time scales, we refer to rapidly adjusting components as "fast variables" and gradually evolving components as "slow variables". The assignment of these fast and slow variables depends on the underlying dynamics. Tracking the slow manifold geometry requires fast variables, whereas only the slow variables are often apparent as the observed dynamics. The essential idea of using HASC as an EWS for B-tipping is to analyze all the variables directly to capture this geometry. Further discussions on this theoretical background are presented below and in the Appendix A.

As described in Section 2, HASC quantifies the instantaneous distribution of effective diffusion-like modes on the data manifold; higher entropy indicates a more isotropic and multi-directional spreading of trajectories. Therefore, in the context of tipping analysis, HASC is expected to be higher when: (i) the quasi-stable attractor is chaotic; (ii) data points around the trajectory are spreading; or in some trivial regimes (iii) the dynamics become excessively slow, and the trajectory flattens (as anomaly; see Appendix B). Conversely, HASC should be lower when: (i) the quasi-stable attractor is non-chaotic or directional; or (ii) the trajectory collapses near a bifurcation point. These characteristics can be exploited to detect tipping phenomena ongoing in high dimensional systems.

Based on these theoretical preparations, we structure the following validation. First, we focus on B-tipping (Section 4.1.), which is the most classical scenario for critical transitions in the climate system. We validate the indicator using both a theoretical low-dimensional model and a high-dimensional Earth-system simulation, highlighting its potential of practicality and how system dimensionality and noise levels affect the HASC signal. We then discuss the applicability of HASC to N-tipping (Section 4.2.) and R-tipping (Section 4.3.) using theoretical models. As explained in Section 2.4., throughout this section, windows to calculate HASC are purposely right-aligned to have fair discussions as a warning signal.

## 4.1. B-tipping

### 4.1.1. Low-dimensional model

In order to evaluate the performance of HASC indicator as EWS, we first analyzed the Atlantic Meridional Overturning Circulation (AMOC) five-box model and its output by Zimmerman et al. (2025). This model is composed of 10 ODEs, and full formulation of this model can be found in (Wood et al., 2019). Zimmerman et al. (2025) calibrated the model and compared the analysis results of conventional EWS indicators based on Critical Slow Down (CSD), Auto correlation (AC1) and Variance (Var), in tipping scenario ($K_N = 5.456\ S_v$, $K_S = 5.447\ S_v$, bi-stable) with non-tipping scenario ($K_N = K_S = 27\ S_v$, monostable) (see the original paper for the details of the model parameters). Interestingly, the result shows that the conventional EWS indicators behave similarly in both tipping and non-tipping scenario. They have successfully raised a crucial example where conventional EWS indicators falsely work.

To follow the discussions by Zimmerman et al. (2025), we adopt three-box reduction first introduced by Alkhayuon et al. (2019) instead of the full five-box model. In this reduction, among original five boxes, North Atlantic (N), Atlantic thermocline (T), Indo-Pacific thermocline (IP), Deep water (D) and Southern Ocean (S), salinities in S and D are assumed to be constant. This simplified model exhibits a similar dynamical structure to the original model for B-tipping. In the following discussions, we adopt the notation of $S_X$ for describing the salinity in box "X", and $T_X$ for the temperature in box "X". The AMOC simulation is carried out by linearly varying the freshwater hosing from $-0.5 Sv$ to $0.5 Sv$ with additive Gaussian noise with intensity $\sigma = 0.09$ and $\sigma = 0.9$.

Conventional EWS indicators are AC1 and Var of the strength of the thermohaline overturning Q

calculated based on a method referred to in (Zimmerman et al., 2025). For HASC, the input for each timestep is $(Q, T_N - T_S, S_N - S_S)$. This vector includes fast variable $(T_N - T_S)$, slow variable $(S_N - S_S)$ and diagnostic state $(Q)$. All indicators are calculated until the first realization tips ($Q < 0$, using right-aligned window). The analysis is conducted over 100 representatives run by noisy hosing. Note that HASC is not a moment-based statistical indicator, therefore it is calculated for each realization.

Unlike the case of Lorenz models in Section3, these variables have physical meanings, thus, they have different scaling. Therefore, we apply the standardization process before calculating HASC value using only the data from 1.5kyr~2.5kyr data. This fixed standardization window setting is adopted to prevent the use of future data for warning signal calculation while avoiding the first spin-up phase.

We first focus on the low-noise regime ($\sigma = 0.09$) to isolate the predominantly bifurcation-induced tipping phenomena. For this analysis, we employ a relatively short time window of W=200 ($HASC(200,50,15)$). While larger windows are typically preferred in high-noise regimes to smooth out stochastic fluctuations, it is expected that the deterministic trajectory dominates the result in this regime. Under this assumption, a shorter window allows us to capture the instantaneous signal without smoothing effect, which is crucial for understanding the quasi-static evolution as it approaches the bifurcation. Figure 6 presents the comparison between conventional indicators and HASC in this regime. As shown in the upper subplots, AC1 exhibits an increasing trend both in tipping and non-tipping scenarios, eventually saturating near 1.0. Variance and VNE, on the other hand, demonstrate specificity. In the non-tipping scenario (Panel (b)), both remain stable throughout the simulation, while in the tipping scenario (Panel (a)), both show imminent reaction before the transition (0.1-0.2$Sv$).

In the high-noise regime ($\sigma = 0.9$), which is the same setting in (Zimmerman et al., 2025), on the other hand, the result is not so simple. Note that for stabilization of statistical estimation, we adopt the window size of W=2000, following the same literature. Figure 7 shows that the conventional EWS indicators exhibit evident increasing trends in both tipping and non-tipping scenarios while HASC indicator remains largely flat in both scenarios. This result for conventional EWS indicators reproduces that of the original paper. Shifting our focus to HASC, although it avoids generating the false positives, it also fails to provide a precursor in this setup. Note that HASC shows the same behavior with a shorter window W=200 (See Supplement S2), which implies that this result is not caused by a mere obscuration by the long averaging window.

However, with further an event-aligned analysis, we find the same reaction with the low-noise regime against tipping in HASC (Figure 8(c)). With a short window aligned to tipping time (W=200), HASC exhibits the same drop with the low-noise regime (Figure 8(a)) imminently (≈200yr) before the tipping. This reaction is imminent, thus, is smeared out in a longer window (W=2000, Figure 8(d)). This result suggests that the drop in HASC analysis is not an EWS indicator, but a conditioned precursor characterized by the distinctive reduction in the system's effective dimension (See N-tipping section for more detailed discussions on the conditioned precursor).

This seems to imply that HASC measures fundamentally different phase compared to the conventional indicators. Conventional EWS indicators are designed to measure the slowly ongoing loss of stability in slow manifold-constrained evolution, whereas HASC measures the actual slow manifold collapse closer to the transition.

One possible reason why HASC fails to capture this early deformation is that the 3box AMOC model has only limited degrees of freedom and the system lacks the spatial dimension to explore even if the dynamical confinement to slow manifold weakens. Therefore, we hypothesize that in a high-dimensional system, this gradual dynamical deformation will emerge as an inflation of effective dimension long before the collapse. Based on this hypothesis, we now turn to high-dimensional earth scale simulation data. This analysis serves not only to validate this hypothesis but also to demonstrate the practical utility of HASC in a realistic setting.

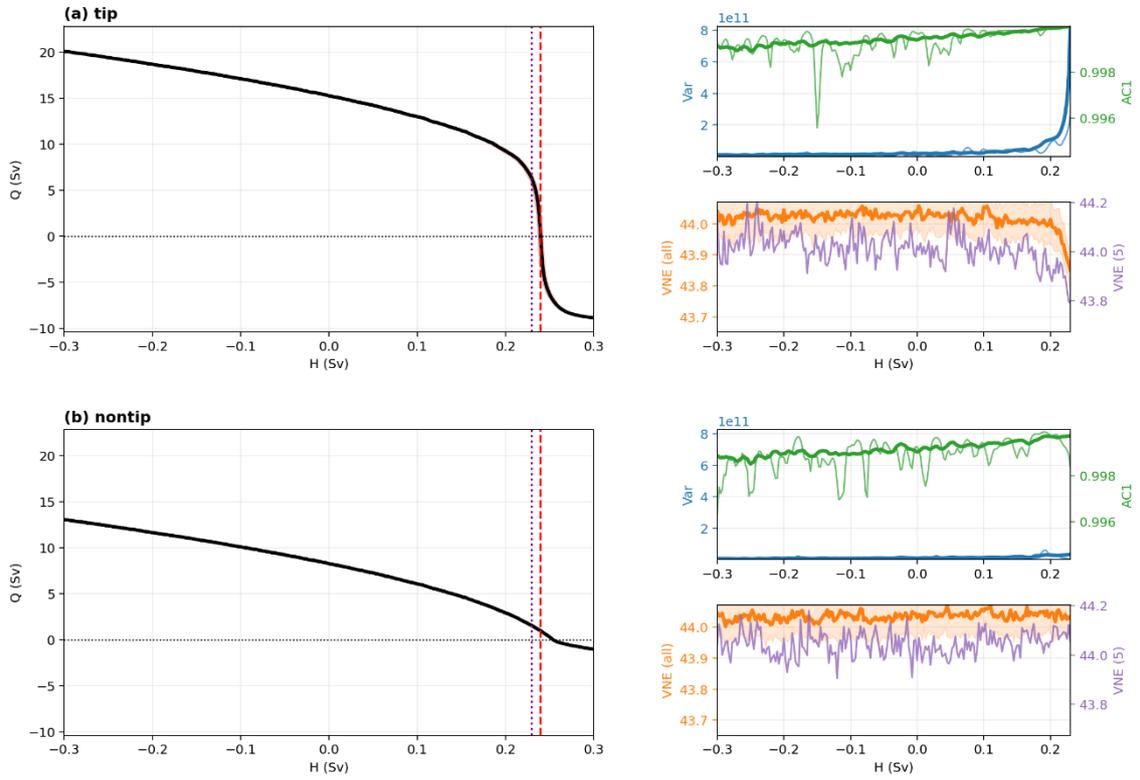

**Figure 6.** Comparison of indicators in the 3box AMOC model under the low noise regime ($\sigma = 0.09$) with a short window (W=200). The time evolution of indicators in shown for (a) tipping and (b) non-tipping scenario. Left panels: The simulation trajectory (solid line) is plotted. The vertical purple dashed line indicates the time first realization crosses Q=0, and red one indicates the tipping time. Right panels: The upper plots show the AC1 (green, right axis) and Variance (blue, left axis), calculated using all the ensemble (solid line) and only 5 realizations (thin line). The lower plots show the HASC indicator (VNE), with the ensemble median and shaded IQR in orange and 5 realizations median in purple line.

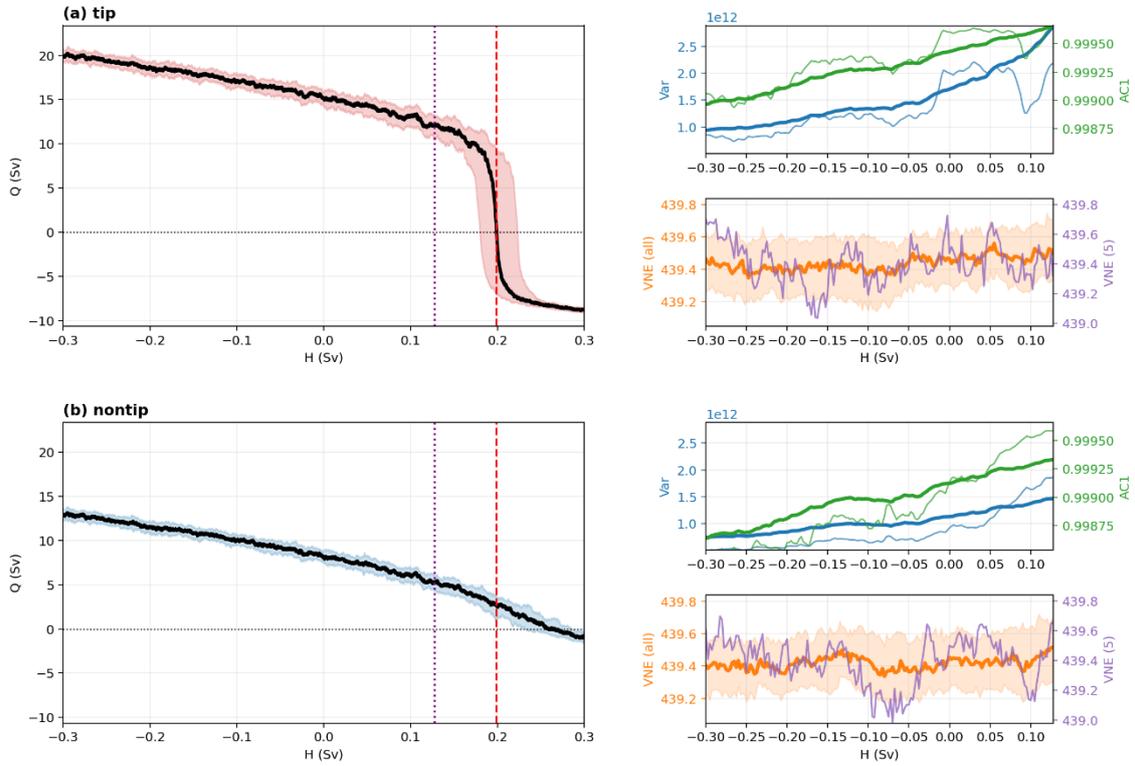

Figure 7. Comparison of indicators in the 3box AMOC model under the high-noise regime ($\sigma = 0.9$) with a long window (W=2000). The layout follows Figure 6. Left panels: The simulation trajectory (solid line) and IQR (shading) are plotted. The vertical purple dashed line indicates the time first realization crosses Q=0, and red one indicates the tipping time. Right panels: The upper plots show the AC1 (green, right axis) and Variance (blue, left axis), calculated using all the ensemble (solid line) and only 5 realizations (thin line). The lower plots show the HASC indicator (VNE), with the ensemble median and shaded IQR in orange and 5 realizations median in purple line.

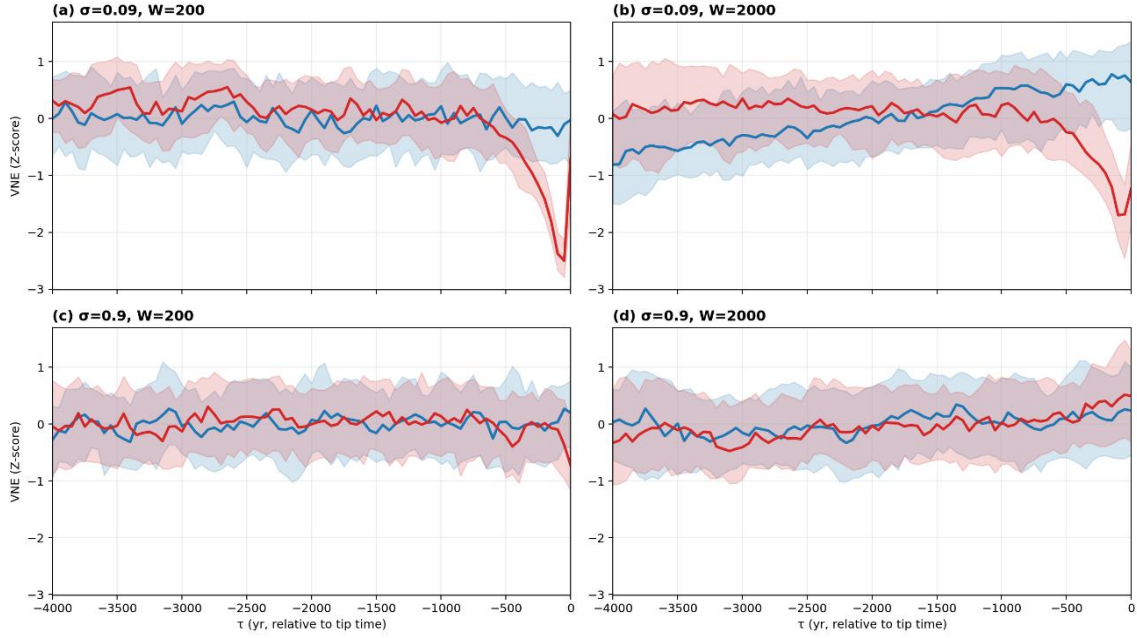

**Figure 8. Event-aligned evolution of the HASC indicator (VNE) across four noise and window conditions.** The VNE time-series are Z-score normalized per run and aligned relative to the tipping time ($\tau = 0$). For non-tipping runs, alignment is based on the median tipping time of the tipping ensemble. The red lines and shaded areas represent the median and IQR of the tipping ensemble, while the blue lines and shaded areas represent the non-tipping ensemble. (a) Low noise ($\sigma = 0.09$) with a short window (W=200). (b) Low noise ($\sigma = 0.09$) with a long window (W=2000). (c) High noise ($\sigma = 0.9$) with a short window (W=200). (d) High noise ($\sigma = 0.9$) with a long window (W=2000).

### 4.1.2. High-dimensional simulation data

In this section, we use the publicly available output of the Community Earth System Model (CESM) State-of-the-Art freshwater-hosing experiment that induces a B-tipping AMOC collapse, as reported by van Westen et al. (2024). The experiment was branched from the end of a preindustrial control simulation for 2800yrs, and external forcings were kept fixed at preindustrial levels. A slowly varying surface freshwater flux anomaly was applied over the North Atlantic and was linearly increased at $3 \times 10^{-4} Sv\, yr^{-1}$ up to $0.66 Sv$ by model year 2200. In this simulation, AMOC strength is defined as the meridional volume transport at $26°N$ integrated over the upper 1000m, and the AMOC tipping point is reported at model year 1758 based on break regression analysis. The processed model output and accompanying scripts are archived on Zenodo(DOI:10.5281/zenodo.10461549) (van Westen, 2024). The analysis using the conventional EWS indicators is reported as "unreliable" by van Westen et al. (2024), and physics-based early warning signal $F_{ovS}$ is discussed as a reliable indicator in this literature.

We integrate slow and fast variables as the input of HASC analysis to test our hypothesis in the last section; loss of stability in a high-dimensional system appears as an inflation of effective dimension in trajectory geometry. As an empirical proxy data, we use (i) the Atlantic meridional overturning stream function field $\Psi(y, z, t)$ (denoted as "AMOC structure", depth×latitude) as a slowly evolving field, and (ii) the Sea surface temperature ("SST", latitude×longitude) field over the Atlantic as a fast-sensitive field. These fields are standardized and concatenated to form a high-dimensional state vector. Standardization statistics are computed from an early pre-tipping baseline period (first 100 years) and applied to the full record.

Specifically, to ensure that the detected signals are robust to the method of state vector construction and to evaluate the contribution of fast variables, we compare three different preprocessing schemes: (i) Raw (AMOC structure 18910 dimensions/SST 13920 dimensions) (ii) Interp (interpolation, SST grid resampled to 18759 dimensions for a balanced contribution) (iii) Lat profile (latitude-wise flattened, AMOC structure data averaged over depth 310 dimensions/SST field averaged zonally and interpolated to the AMOC latitude grid). We also carry out a control experiment using only the AMOC structure data (18910 dimensions), excluding SST. We analyze the output data with $HASC(300,1,15)$ based on the insights from the 3box model analysis, where shorter windows are advantageous to capture local quasi-stable geometry.

Figure 9 presents the results of the HASC analysis. The top panel shows the AMOC index (blue) and the physics-based early warning signal $F_{ovS}$ (freshwater transport at $34°S$, orange), whose minimum at model year 1732 serves as a precursor (van Westen et al., 2024). Focusing on the HASC results in the bottom panel, in contrast to the 3box model results, the CESM analysis with fast variables (all the lines except for the red one) reveals a distinct precursory rise in HASC starting about 1200 model yrs before the tipping event. The increase is particularly evident in the "interp" scheme (green line). This supports our hypothesis in the last sections and visualizes that SST as fast variables are allowed to explore a wider phase space with gradual loss of effective confinement to low-dimensional slow manifold-like backbone. Following this increase, HASC peaks around model year 1600 and drops, preserving the imminent reaction to tipping discussed in the last section. This timing is closely related to the timing when $F_{ovS}$ approaching the bottom and start to fluctuate.

Figure 10 further confirms the robustness of this detection. With "interp" scheme, across a range of window sizes (W=100-400), the "rise and drop" pattern is consistent. This result highlights the critical advantage of HASC in high-dimensional real-world data: instead of focusing on low-dimensional projections, the full state space analysis captures the tiny system destabilization as a measurable geometric signal.

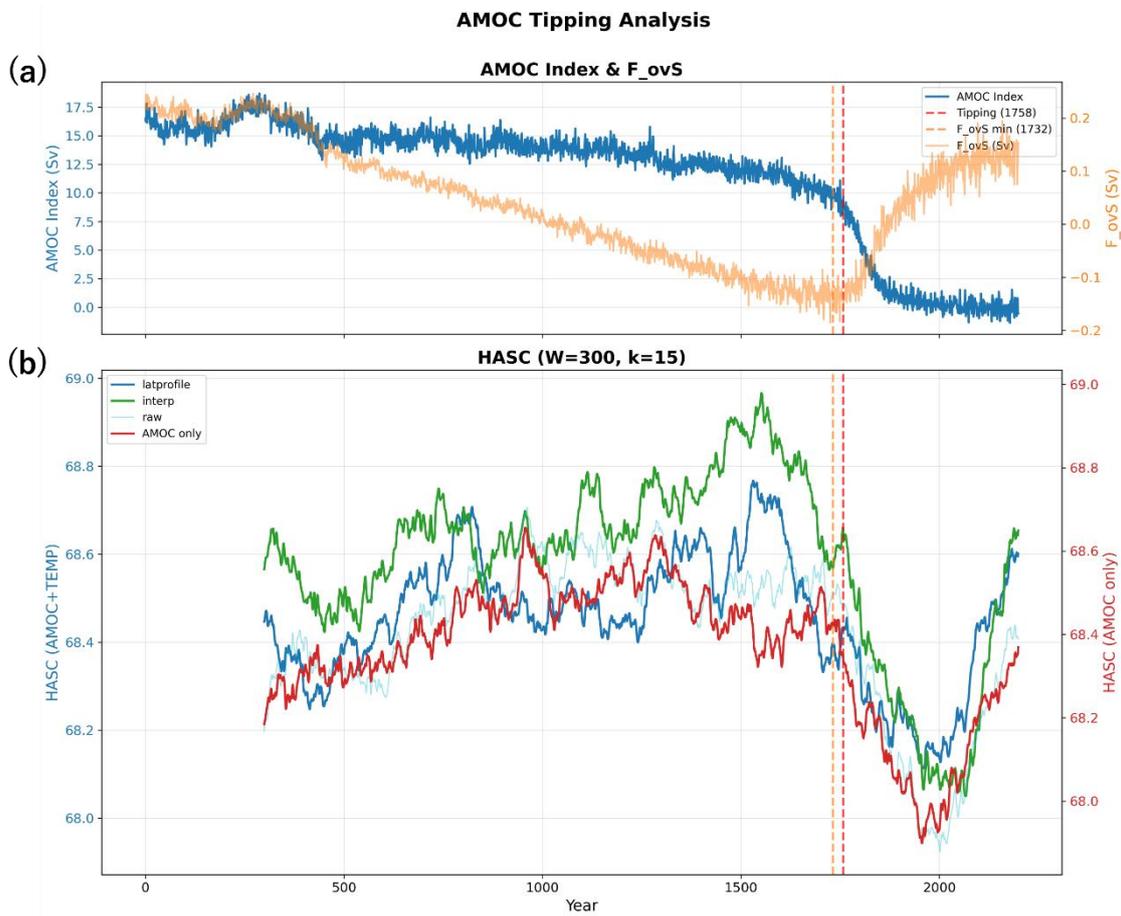

**Figure 9. Application of HASC to the CESM freshwater-hosing simulation. (a) Time evolution of the AMOC strength index (blue, left axis) and the freshwater transport at $34°S$ ($F_{ovS}$, orange, right axis). The vertical dashed lines indicate the minimum of $F_{ovS}$ at model year 1732 and the tipping point determined by break regression at model year 1758. (b) Time evolution of the HASC indicator comparing three state-space construction schemes and a**

control experiment: The AMOC structure only (red), "Raw" (AMOC structure+SST, light blue), "Interp" (AMOC structure+interpolated SST, green) and "Lat profile" (AMOC structure averaged for depth+interpolated zone averaged SST, blue).

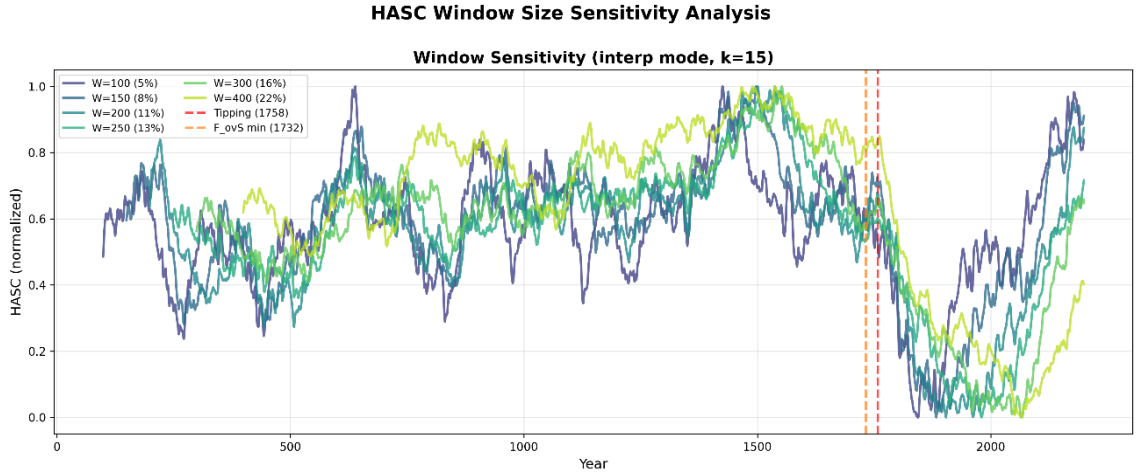

**Figure 10. Robustness of the HASC signal against window size variations with "Interp" scheme. The time evolution of the HASC indicator is shown for window sizes ranging from W=100 to 300 years, The values are min-max normalized for comparison.**

## 4.2. N-tipping

We utilized the same model as B-tipping 3box model experiment with the hosing term $H = 0.2 + \sigma \cdot \varepsilon(t), \sigma = 0.14$ with $\varepsilon(t) \sim \mathcal{N}(0,1)$, i.i.d. to isolate predominantly N-tipping phenomena. We generated an ensemble of 250 realizations, filtering for "genuine" N-tipping events that occur after t=4.0kyr to exclude tipping (judge by $Q < 0$ for more than 50yrs) by initial transient instabilities. For the HASC analysis, the setting is $HASC(1000,50,15)$ with the input vector $(Q, T_N - T_S, S_N - S_S)$. We only employ a shorter window length, as long-term EWS reactions are generally absent in N-tipping (Ditlevsen & Johnsen, 2010). To reveal systematic structural changes despite the stochastic timing of tipping, we aligned the HASC time series of tipped realizations to their individual tipping onset time, defined by the crossing of $Q = 0$. A standardization process was applied relative to a distant pre-tipping baseline period -4.0kyr~-3.0kyr for each realization, thereby normalizing the varying magnitudes of the state variables and ensuring that the comparison captures relative deviations from the stable state.

As shown in Figure 11(a), the ensemble clearly separates into tipped (red) and non-tipped (blue)

trajectories driven by the stochastic forcing. The corresponding HASC analysis (Figure 11(b)) reveals a distinct precursor signature. While the non-tipped reference trajectories maintain a stable entropy level, indicating isotropic diffusion within the basin of attraction, the tipped ensemble exhibits a sharp and systematic decrease in median VNE approaching the tipping point ($\tau = 0$).

Quantitative evaluation of Cliff's delta (Figure 12(a)) confirms the statistical significance of this signal. The effect size drops below zero, crossing the effective threshold approximately 100 to 200 years prior to the tipping event. This negative $\delta$ confirms that the entropy of tipped trajectories is stochastically dominated by the non-tipped reference. The robustness of this detection is supported by the sensitivity analysis (W=200-2000) in Figure 12(b), where the decreasing trend remains consistent across a wide range of window sizes (W=200-800). However, for longer window sizes, the signal is smoothed and thus becomes unobservable.

This behavior can be explained from insights from transition-path theory. By aligning the data to the tipping time, we effectively condition the ensemble on the occurrence of the event, isolating reactive trajectories from the general trajectories. Reactive trajectories do not explore the phase space randomly; instead, they concentrate into narrow transition channels as they cross the effective barrier in the quasi-potential landscape (E. & Vanden-Eijnden, 2006). This concentration implies a reduction in the effective dimensionality and directional freedom of the trajectories, which HASC detects as a decrease in entropy. Recent AMOC studies explicitly utilize rare-event sampling algorithms (e.g., Trajectory-Adaptive Multilevel Splitting (TAMS)/Adaptive Multilevel Splitting (AMS)) to efficiently generate and analyze these rare transition paths (Castellana et al., 2019; Jacques-Dumas et al., 2024). These methods leverage the statistical concentration of trajectories along preferred transition pathways to estimate tipping probabilities. From this standpoint, the HASC signal serves as a geometric imminent EWS: it captures the local reorganization and confinement of trajectory geometry as the system enters the transition channel, which becomes visible through this conditional experiment. Indeed, the concept of utilizing trajectory geometry as an EWS has recently gained attention. For example, Chapman et al. (2025) explicitly proposed geometric early warnings based on the quasi-potential landscape for this same AMOC model, reinforcing the validity of geometric approaches in N-tipping detection.

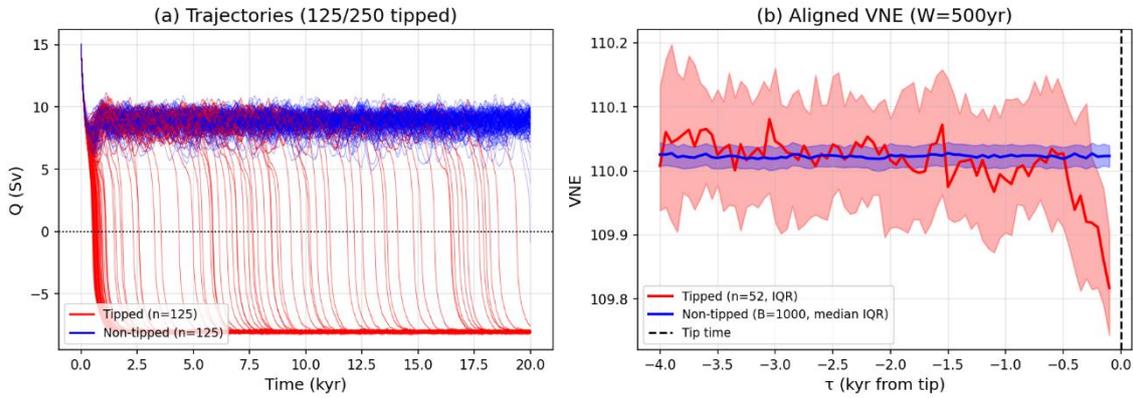

**Figure 11. Noise-induced tipping dynamics and HASC analysis.** (a) Ensembles trajectories of the AMOC strength Q (250 realizations). The system is driven by a fixed mean forcing $0.2 Sv$ with additive Gaussian noise ($\sigma = 0.14 Sv$). Red curves indicate tipped realizations which satisfy the persistence criterion (Q<0 for more than 50 consecutive years), while blue curves indicate non-tipped ones. (b) Aligned time evolution of HASC time-series. Tipped realizations are aligned to the tipping onset time. Solid lines denote the ensemble medians, and shaded bands represent IQR.

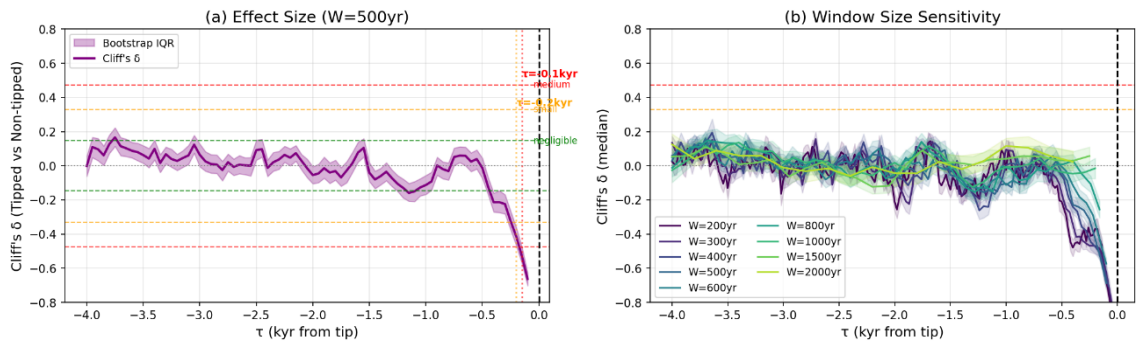

**Figure 12. Statistical quantification of the pre-tipping signal.** (a) Time-dependent effect size (Cliff's delta) comparing the aligned HASC time-series distributions of tipped vs non-tipped realizations (W=500). The solid line indicates the estimated $\delta$, while the shaded band represents the bootstrap IQR. Horizontal colored dotted lines denote the standard effect size thresholds, negligible (green), small (orange), medium (red). Vertical colored dotted lines mark the times when the signal crosses small, medium threshold relative to the tipping onset. (b)

**Window size sensitivity analysis. Solid colored lines correspond to the median $\delta$ calculated for different window lengths ranging from W=200 to 2000yrs.**

## 4.3. R-tipping

To investigate R-tipping, we employed a transient forcing protocol to the same model with hosing term:

$$H = \Delta H \cdot a(r,t), a(r,t) = \begin{cases} \text{sech}\left(r(t - T_{peak})\right), 0 \leq t \leq T_{peak} \\ 1, t > T_{peak} \end{cases} \quad (18)$$

, where $\Delta H$ (total magnitude of the forcing increase) is $0.365Sv$, $r$ (rate parameter) is 0.0177 and $T_{peak}$ (timing of the inflection point) is 1000. This smooth ramp forcing setting in the AMOC box model follows (Ritchie et al., 2023). This profile creates a two-phase forcing scenario: an initial slow-change phase followed by a rapid transition to a new forcing level.

Here we explore the "basin instability" of the system. In multi-stable systems, qualitatively distinct outcomes are separated by a threshold; a separating set associated with the basin boundary between competing attractors. From this viewpoint, R-tipping can be interpreted as the nonautonomous trajectory crossing a time-varying basin boundary (threshold) (Wieczorek et al., 2023). The regime of "basin instability" occurs when the system fails to track the moving base state and is overtaken by the shifting basin boundary. We thus generated a dense grid of 81×81 initial conditions in the scaled state space $(S_N, S_T) \in [0, 0.25]^2$ around the base equilibrium derived from 5000yrs spin-up with no hosing. Each initial condition is subjected to the same forcing protocol. Tipping events are classified using the same persistence criterion as the N-tipping experiment (Q<0 for more than 50 consecutive years).

We analyze the time-series data with $HASC(200,10,15)$ and the input vector to HASC is $(Q, T_N - T_S, S_N - S_S)$. For a standardization, we concatenated the simulation data with a 1000yrs spin-up tail (equilibrium) for each trajectory and apply the standardization process only using the statistics of the spin-up tail to ensure that the VNE calculation captures the deformation relative to the baseline equilibrium.

Figure 13 illustrates the time evolution of the ensemble under forcing explained above. Although all trajectories are subject to the identical forcing profile (Figure 13(a)), the ensemble bifurcates into tipped (red) and non-tipped (blue) groups (Figure 13(b)). The corresponding HASC analysis reveals a distinct time-evolution in structural complexity for each group before the critical transition occurs (Figure 13(c)).

To quantify this separability, we compute a distance score $d = d_{non} - d_{tip}$ based on the pre-peak VNE time-series, where $d_{tip}$ and $d_{non}$ are correlation distances to prototypical tipped and non-tipped patterns, respectively shown in Figure 13(d). The similarity is defined by the correlation distance. Figure 14 compares (a) the true basin of attraction determined by the forward simulation with (b) the basin reconstructed solely from the pre-peak VNE distance score defined above. The reconstruction shows a remarkable agreement with the ground truth, effectively mapping the complex, non-linear separatrix between the survival and tipping regions.

It is widely known that R-tipping is not induced by the system's destabilization, which causes statistical EWS to fail (Ritchie & Sieber, 2016). Our results demonstrate that HASC can serve as a robust geometric indicator of the underlying instability, and in scenarios where the actual collapse is delayed due to the transient nature of the forcing, this geometric indication can act as a precursor if the basin structure is partially known. We can further give compelling interpretations of the VNE evolution (Figure 13(c)). The peak just before the 1000yr (=$T_{peak}$) in the tipping ensemble corresponds to the phase of rapid geometric deformation as they approach the "edge state"; a saddle-like invariant set that organizes the basin boundary structure, especially for this model, identified as a repelling periodic orbit rather than a simple fixed saddle point (Ritchie et al., 2023). The drop and stabilization after this suggest that the ensemble aligns along the unstable manifold of the orbit, resulting in the temporal effective dimension reduction before the collapse.

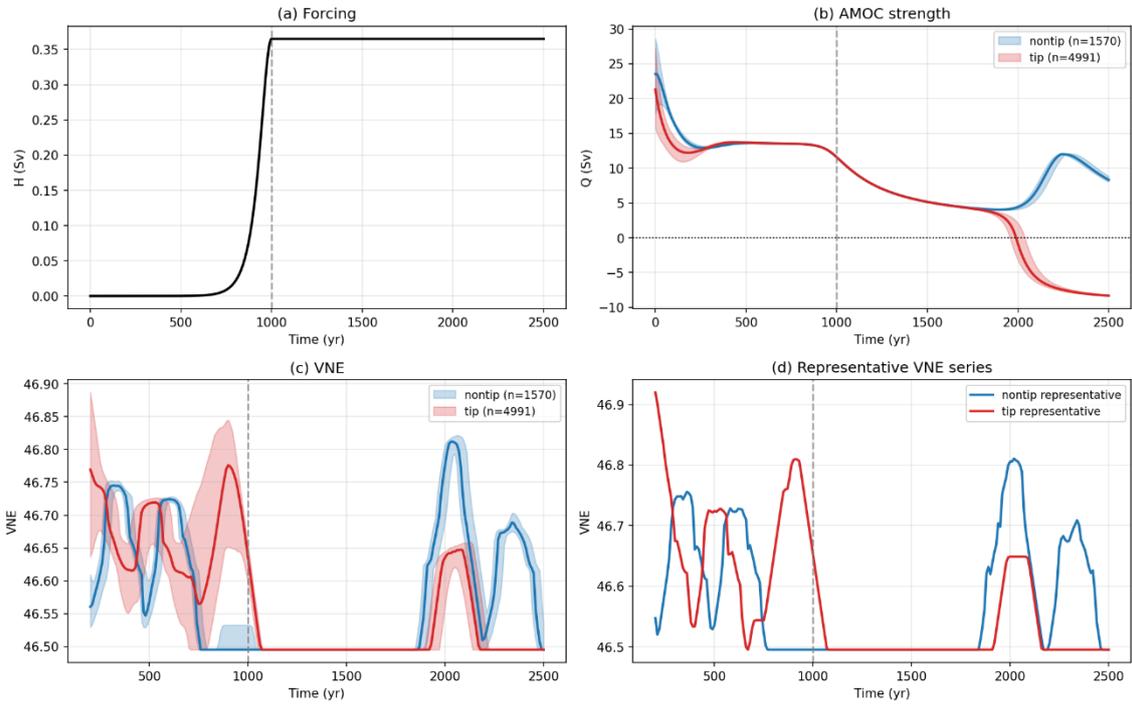

**Figure 13. R-tipping dynamics under transient ramp forcing. (a)** Time evolution of the freshwater forcing, following a hyperbolic tangent profile (Δ*H*=0.365, r=0.0177). **(b)** Ensemble trajectories of AMOC strength Q. Red: tipped; Blue: non-tipped. **(c)** Time evolution of the VNE for the ensemble HASC analysis. Shaded bands represent the IQR. **(d)** Representative single VNE time-series for a tipped (red) and non-tipped (blue) trajectory.

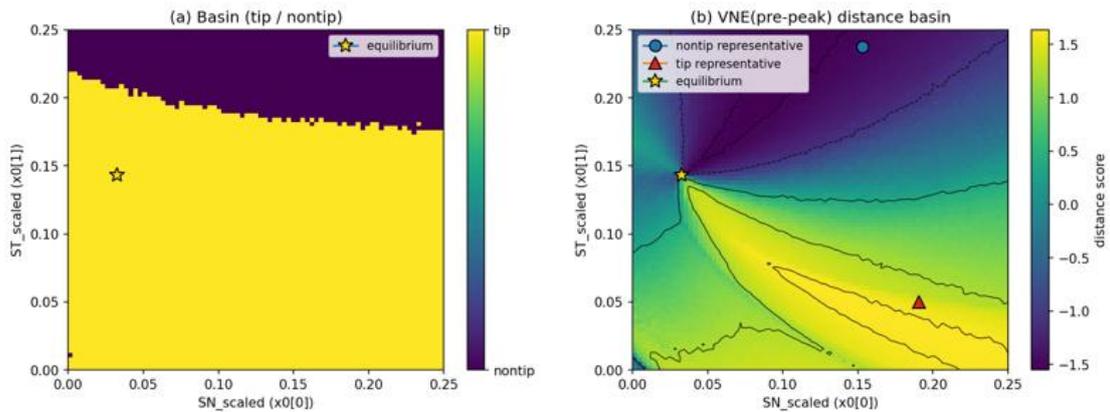

**Figure 14. Geometric reconstruction of the basin of attraction. (a)** True basin of attraction map in the scaled initial condition space $(S_N, S_T) \in [0, 0.25]^2$. Yellow pixels indicate

**tipping, while purple pixels indicate survival (non-tip). The star denotes the equilibrium point of the spin-up. (b) Reconstructed basin map based on the VNE distance score, calculated using the correlation distance between the pre-peak VNE time-series and the class prototypes. The color scale represents the score, where higher values correspond to tip-like and lower values correspond to non-tip-like geometry. The contours in the basin are plotted based on the 5, 25, 50, 75, 95th percentiles of the calculated score distribution.**

# 5.Conclusion

Detecting and forecasting tipping phenomena in complex nonlinear systems are challenging tasks especially when the effective state is high-dimensional, observations are partial and noisy, and the system is non-autonomous. In such regimes, classical EWS based on CSD or low-order statistics are not always reliable, and purely dynamical indicators are practically difficult to apply.

To address this gap, we propose a unified pipeline that couples state reconstruction via DA with a measure of a geometry-aware representation of windowed trajectory data and a spectral information complexity. This measure is derived from the graph Laplacian obtained through trajectory approximation. Here the indicator is referred to as HASC, and the entire pipeline as DA-HASC.

The target of HASC is distinct from simple "chaoticity"; instead, it quantifies the effective degrees of freedom in high-dimensional dynamics. Therefore, it is designed to capture changes in attractor organization, transition pathways, and local connectivity patterns. In this sense, across B-, N-, and R-tipping, the same geometric quantity exhibits mechanism-specific signatures: precursory inflation and following collapse in B-tipping, event conditioned channelization in N-tipping, and deformation of basin-geometry in R-tipping. Additionally, we show that it is applicable to high-dimensional (more than $10^4$) climatological simulation data and can even extract more information compared to low-dimensional model. Thus, as an indicator, HASC is not what is conventionally called EWS, but a unified structural indicator whose lead time is mechanism- and dimension-dependent.

A key next step is to establish more systematized hyper-parameter sensitivity, state vector preprocessing, and standardization protocols to clarify trade-offs between lead time, robustness, and false positives. This requires deep insights on scale-/feature-selection in dynamical systems and left for future work. At the same time, consistency and finite-sample error analysis for Laplacian spectra and VNE approximation are necessary to further justify high-dimensional applications. For improved applicability, scalability and reproducibility, incorporating density-aware manifold construction (DensMAP; (Narayan et al., 2021)) or better approximation method of VNE for large graphs (e.g. (Benzi et al., 2023)) is also important. Of course, further validations and extensions especially for N-tipping, R-tipping and real-world data are promising areas.

Overall, this work positions HASC as a geometry- and spectrum-based structural indicator that can bridge simulations or reanalysis data and nonlinear time-series analysis, offering a practical route to characterize and monitor regime change in non-autonomous high-dimensional dynamical systems under realistic constraints in observation and computation.

## Supplementary material

See the supplementary material for additional figures(S1/S2).

## Author declarations

### Conflict of Interest

The authors have no conflicts to disclose.

## Data availability

The code and processed data generated in this study will be made publicly available via Gitlab following an embargo from the date of publication to allow for curation. The external data analyzed in this study are available from van Westen (2024) via Zenodo (see References).

### Appendix A: Theoretical backgrounds of HASC as tipping analysis indicator

The idea of using HASC as an EWS is rooted in Geometric Singular Perturbation Theory (GSPT), which introduces the concept of slow manifolds in multi-scale systems and provides a rigorous definition on transition points. It is well established that when multi-scale systems have explicit timescale separation in their subsystems, one can effectively model the critical transition using a fast-slow system, typically expressed as:

$$\varepsilon \frac{dx}{d\tau} = \varepsilon \dot{x} = f(x, y) \quad (A1)$$

$$\frac{dy}{d\tau} = \dot{y} = g(x, y) \quad (A2)$$

where $\tau = \varepsilon t, 0 < \varepsilon \ll 1$. Considering the singular limit $\varepsilon \to 0$ yields the slow subsystem, whose flow is constrained to $C = \{(x, y) \in \mathbb{R}^{m+n} : f(x, y) = 0\}$. The set is called the critical manifold if $C$ has manifold structure. For any normally hyperbolic submanifold $S_0 \subset C$, Fenichel's theorem stands: there exists a locally invariant manifold $S_\varepsilon$ diffeomorphic to $S_0$ for sufficiently small $\varepsilon > 0$, and the flow on $S_\varepsilon$ converges to the slow flow as $\varepsilon \to 0$. $S_\varepsilon$ is called a slow manifold and usually is not unique. However, a family of slow manifolds lies within an exponentially small distance and choice of representative is negligible for the asymptotic analysis; thus, we can call "the" slow manifold here (Kuehn, 2013).

B-tipping occurs when the slow manifold itself loses stability (i.e., loss of normal hyperbolicity) via certain classes of bifurcation, a typical example is saddle-node (fold) bifurcation. In this process, based on Critical Slowing Down (CSD) concept, which claims that dynamical systems respond slowly to perturbations as they approach a tipping point (Scheffer et al., 2009), conventional EWS such as Auto correlation (AC1) or Variance (Var) works. This is because, with loss of normal hyperbolicity, eigenvalues approach zero and center direction emerges locally around the transition point (Guckenheimer & Holmes, 1984).

On the other hand, when we introduce stochastic perspective in this theoretical framework, it reveals a strip around the deterministic slow manifold. N-tipping occurs when orbits exit this stable strip due to strong stochastic perturbations (Kuehn, 2011), without the collapse of the slow manifold itself. R-tipping occurs when $\varepsilon$ is not sufficiently small anymore and a trajectory fails to track the slow manifold, which, again, is not induced by the manifold collapse (Wieczorek et al., 2023). Therefore, the effectiveness of EWS for N-tipping and R-tipping is questionable.

From this theoretical perspective, as HASC effectively reconstructs and evaluates the local geometry of the underlying invariant manifold, it is expected that its value reflects structural changes during the tipping phenomena. In fact, as discussed by Chapman et al. (2025), utilizing local geometrical changes in slow manifold as Geometrical EWS is being studied in AMOC box model tipping simulation induced by transient forcing. Therefore, at least for B-tipping, detection of CSD and degeneracy to central flow by HASC should work as an EWS-like signal. For N-tipping and R-tipping, HASC value would work as very imminent EWS or just for detection, as there exists no collapse of slow manifold. The discussions on this can be found in N-tipping and R-tipping section. Note that in order to correctly evaluate the input with HASC, it must contain the slow variables one wants to focus on, and high-dimensional spatial or multivariate time-series to prevent topological degeneracy. When inputting different physical units and scales, each variable should be standardized to ensure equal contribution to the structure.

## Appendix B: Anomaly of HASC against converging trajectory

Here we show the numerical anomaly of HASC calculated against strongly converging trajectory, caused by k-NN graph construction algorithm.

In Figure B1, the attractor is structurally less complex (convergence) and LLE stays around 0. On the other hand, HASC shows rapid increase after the spin-up phase. This behavior is triggered by an indistinguishably small variance in the k-th nearest neighbor distances within each window.

More precisely, for each point $x_i$ in a window, let $d_i$ denote its distance to its k-nearest neighbor. When a window contains only either a truly uniform point cloud or a locally collapsed nearly point-like segment of the trajectory, the standard deviation:

$$\sigma_k = \sqrt{\frac{1}{N}\sum_{i=1}^{N}(d_i - \bar{d})^2} \qquad (B1)$$

approaches 0. This poses a challenge for the k-NNG construction in HASC based on UMAP. UMAP constructs a graph by estimating a characteristic local scale for each point derived from k-distances to normalize edge weights (McInnes et al., 2018). Consequently, it cannot distinguish between a truly uniformly distributed point cloud and a locally collapsed point-like segment, since in both scenarios, the relative connectivity resembles that of a complete/regular graph where every node is equally connected to others. This results in the HASC value suddenly starting to rise and staying at the maximum value when the points inside a window exhibit point-like concentration, as VNE is maximized when the graph spectrum is uniform (=complete/regular graph). We refer to windows showing this phenomenon simply as a "locally uniform" phase. This term includes uniform point cloud from k-NN perspective and pseudo-uniform point cloud caused by limitations on numerical calculations.

Here, the orbit rapidly converges to a point, naturally resulting in nearly point-like segment and thus extremely small k-distance variance. Indeed, we can observe the "locally uniform" anomaly where HASC value starts to drop then suddenly increase while the attractor structure continues to shrink. Focusing at std(k-dist) plot, this anomaly appears as $\sigma_k < 10^{-7}$, and std(k-dist) stops decreasing after reaching this threshold. This threshold corresponds to the time estimated LLE starts to exhibit huge variance, which proves the point that this anomaly is essentially rooted in k-distance calculation.

Note that "locally uniform" anomaly does not exclusively refer to point-like convergence. The same behavior will be expected with extremely strong sticking to a trajectory or locally collapsed low-dimensional manifold. Although emergence of complete structural uniformity can also cause this anomaly, in real-world physical systems, it is almost negligible due to inherent dynamical constraints and correlations. Thus, HASC value is expected to reflect meaningful shifts below the maximum value quantifying how chaotic the system is, except for the case we've considered above. However, we have to carefully inspect the std(k-dist) value in case of low-dimensional and significantly static theoretical models.

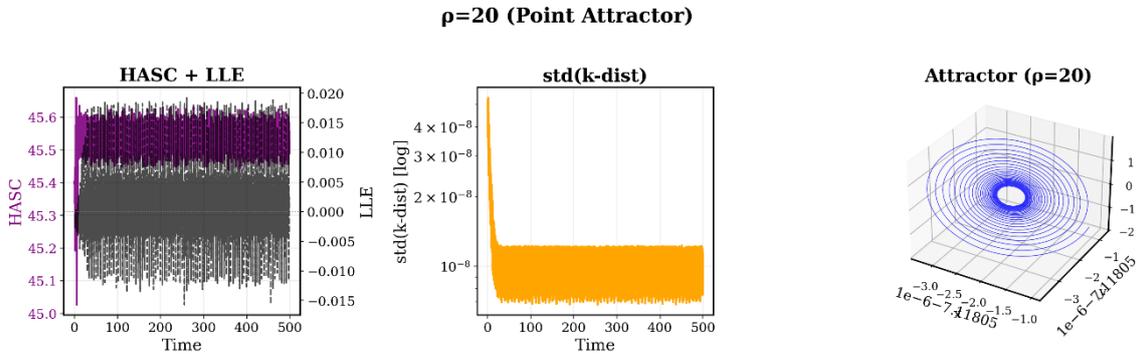

**Figure B1. Plot of HASC/Instantaneous LLE, corresponding standard deviations of kNN distance in a window and the attractor for Lorenz63 model with parameter $\rho = 20$. In the HASC+LLE figure, purple lines shows the HASC value and black lines shows the estimated instantaneous LLE value.**